\documentclass[twocolumn]{openjournal}

\usepackage{natbib}
\usepackage{amsmath,amssymb,verbatim}
\usepackage{graphicx}
\usepackage{hyperref}
\usepackage{txfonts}

\def\ero{eROSITA}

\begin{document}

\title{Low-scatter galaxy cluster mass proxies for the \ero\ all-sky survey}

\author{Dominique Eckert\altaffilmark{1}, Alexis Finoguenov\altaffilmark{2}, Vittorio Ghirardini\altaffilmark{3}, Sebastian Grandis\altaffilmark{4,5}, Florian K\"afer\altaffilmark{3}, Jeremy S. Sanders\altaffilmark{3}, Miriam Ramos-Ceja\altaffilmark{3}
}
\affil{$^1$Department of Astronomy, University of Geneva, Ch. d'Ecogia 16, CH-1290 Versoix
	\email{Dominique.Eckert@unige.ch}}
\affil{$^2$Department of Physics, University of Helsinki, PO Box 64, 00014, Helsinki, Finland}
\affil{$^3$Max-Planck-Institut f\"{u}r extraterrestrische Physik, Giessenbachstrasse 1, 85748 Garching, Germany}
\affil{$^4$Faculty of Physics, Ludwig-Maximilians-Universit\"at, Scheinerstr. 1, 81679, Munich,  Germany}
\affil{$^5$Excellence Cluster Universe, Boltzmannstr. 2, 85748, Garching, Germany}

\begin{abstract}
The on-going X-ray all-sky survey with the \emph{eROSITA} instrument will yield large galaxy cluster samples, which will bring strong constraints on cosmological parameters. In particular, the survey holds great promise to investigate the tension between CMB and low-redshift measurements. The current bottleneck preventing the full exploitation of the survey data is the systematics associated with the relation between survey observable and halo mass. Numerous recent studies have shown that gas mass and core-excised X-ray luminosity exhibit very low scatter at fixed mass. We propose a new method to reconstruct these quantities from low photon count data and validate the method using extensive \ero-like simulations. We find that even near the detection threshold of $\sim50$ counts the core-excised luminosity and the gas mass can be recovered with 20-30\% precision, which is substantially less than the scatter of the full integrated X-ray luminosity at fixed mass. When combined with an accurate calibration of the absolute mass scale (e.g. through weak gravitational lensing), our technique reduces the systematics on cosmological parameters induced by the mass calibration.
\end{abstract}

\maketitle

\section{Introduction}

Galaxy clusters are the endpoint of the structure formation process and they represent the high-mass tail of the halo mass function. As such, their abundance and mass distribution is a sensitive probe of cosmological parameters \citep[see][for a review]{allen11}. Measurements of the galaxy cluster mass function have been obtained using X-ray surveys \citep{reip,boehringer14,mantz15,schellenberger17,pacaud18}, the Sunyaev-Zeldovich effect \citep[SZ,][]{planck20,planck15_24,act13,benson13,dehaan16,bocquet19}, and optical surveys \citep{rozo10,mcclintock19,costanzi19,finoguenov19}, yielding competitive results on cosmological parameters with samples of a few hundred clusters. Recently, some tension has emerged between cosmological parameters measured from the power spectrum of the cosmic microwave background and from various low-redshift cosmological probes, including, but not limited to, galaxy cluster counts \citep{heymans13,hildebrandt16,riess16,riess19,bonvin17,addison16}. If confirmed at a higher level of significance, this tension is indicative either of new physics or of unknown systematics in the various methods. 

In this respect, the upcoming \ero\ mission has an important role to play. \ero\ \citep{erosita} is a wide-field X-ray instrument on board the \emph{Spektrum Roentgen Gamma} (SRG) spacecraft, which was successfully launched on July 13, 2019. \ero\ features a large collecting area at 1 keV of $\sim2,500$ cm$^2$ (on axis) and an angular resolution of 28 arcsec high-energy width (HEW) in survey mode over a field of view (FOV) of 50 arcmin diameter. Combined, these properties yield a survey speed (\emph{grasp}) of 1050 cm$^2$ deg$^2$, which represents an order of magnitude increase compared to previous X-ray instruments. \ero\ will perform eight all-sky surveys over a period of 4 years \citep{merloni12,clerc18}. The sensitivity of the final \ero~survey will be about 20 times deeper in the [0.5-2] keV band than its predecessor the ROSAT all-sky survey. Upon completion, \ero\ is expected to detect 50,000-100,000 individual clusters \citep{pillepich12,pillepich18,grandis18}, which represents an increase by two orders of magnitude compared to current X-ray and SZ cluster samples. The \ero\ survey thus holds great promise for the future of cosmology.

To maximize the cosmological constraints from the survey, several key points need to be properly understood and modeled, which requires an accurate understanding of the physics governing the intracluster medium (ICM). In particular, efficient mass proxies must be defined from survey observables. An optimal mass proxy should combine two fundamental properties: cheap, and efficient. The proxy should be cheap in the sense that it should be measurable with good precision from the survey data. It should be efficient in the sense that the scaling relation between mass and observable should have a low scatter. At first approximation, the state of the gas in galaxy clusters is determined by the properties of the gravitational potential and the merging history of the host halo, which implies the existence of tight scaling laws between ICM properties and cluster mass \citep[the self-similar model,][]{kaiser86,bryan98}. Thus far, most X-ray surveys made use of the total, integrated X-ray luminosity $L_{X,tot}$ as a tracer of the total mass \citep[e.g.][]{ebeling07,ebeling10,boehringer13,boehringer17,giles16}. While the total X-ray luminosity is easy to recover from the survey data, its use as a cosmological tool is hampered by a large scatter at fixed mass \citep[$\sim 50\%$, e.g.][]{arnaud99,maughan07,pratt09,mantz10,giles16}. 

The large scatter of the $L_{X,tot}-M$ relation is caused by the diversity of observed gas properties in cluster cores \citep{croston08,pratt09,lrm09,e12,ghirardini19}. In the central regions ($R\lesssim0.2R_{500}$), the gas densities are such that baryonic physics strongly affects the gas properties. Deviations from self similarity occur in the presence of non-gravitational physics such as gas cooling and feedback from supernovae and active galactic nuclei (AGN) \citep{tozzi01,borgani04,kravtsov05,nagai+07}. As a result, the density profiles of relaxed, cool-core (CC) clusters show a prominent cusp in their central regions, which causes them to be overluminous at a fixed mass with respect to systems exhibiting flat cores (non-cool-core (NCC) clusters). In addition to the large scatter at fixed mass, recent studies have shown unequivocally that X-ray flux limited samples are biased in favor of CC systems because of their higher luminosity and peaked surface brightness profiles \citep{ccbias1,rossetti16,rossetti17,andrade17}. The CC bias greatly complicates the modeling of the selection function, as the exact fraction of CC clusters is still poorly known \citep{ccbias1}.

Conversely, beyond the central regions ($R\gtrsim0.2R_{500}$) the gas density profiles exhibit a high level of self similarity and the scatter of the scaled profiles reduces to $\sim15\%$ \citep{croston08,e12,ghirardini19}. In this regime, baryonic effects are subdominant with respect to gravitational collapse. The scatter of the scaling relations drastically decreases \citep{mantz10,mantz18,maughan12} and the redshift evolution follows the self-similar expectation \citep{mcdonald17}. On top of that, numerical simulations implementing different subgrid models for baryonic physics make very similar predictions beyond  the central regions \citep{nifty1,nifty2}. Thus, cosmological simulations can be used to create reliable mock catalogues to calibrate the selection function. For all these reasons, X-ray cluster samples based on core-excised quantities are well suited for cosmological studies. 

In this paper, we present a novel method for PSF deconvolution and deprojection. The method is based on the decomposition of the observed brightness profiles into a sparse linear combination of basis functions. The basis functions are convolved with the point spread function (PSF) of the instrument and the model is fitted to the observed count profiles using a Hamiltonian Monte Carlo sampler. Our method represents a substantial step forward over traditional methods in the sense that it allows a large freedom in the reconstructed shape, yet at the same time it imposes physically motivated priors on the parameters, makes full use of the statistical properties of the data, and does not depend on the adopted radial binning. We validate the method using extensive \ero-like simulations and demonstrate that it is able to reconstruct accurately the true core-excised X-ray luminosity down to the detection limit of 50 source counts, independently of the core state. Assuming that our method can be combined with an accurate external calibration of the absolute mass scale, we then discuss the implications of our results for the cosmological constraints to be obtained with \ero. We distribute the code as a freely available, easy-to-use Python package\footnote{\href{https://github.com/domeckert/pyproffit}{https://github.com/domeckert/pyproffit}} to allow for an easy replication of our results.

\section{Generalized linear model with sparsity constraints}
\label{sec:method}

For the purpose of this paper we implement a 1D version of the deprojection and PSF-deconvolution algorithm developed by \citet{diaz17}. Namely, we describe the observed profile as a linear combination of basis functions with an $\ell_1$ sparsity term to avoid amplifying statistical fluctuations and enforce smoothness of the reconstructed profile. Here we describe the various steps of the reconstruction. We assume that the position of the cluster center is known a priori and that a surface brightness profile SB$(r)$ has been extracted with the finest possible radial binning, albeit ensuring a minimum of 1 count per annulus.

\subsection{The model}
\label{sec:model}

For a given 3D profile $\epsilon(r)$, we assume that the true profile can be described as a sparse representation of basis functions $\{\Phi_p\}_{p=1}^P$:

\begin{equation}\epsilon(r)=\sum_{p=1}^P\alpha_p\Phi_p(R),\label{eq:3D}\end{equation}

with $P$ the total number of basis functions used, $R$ the 3D distance to the cluster center, and $\alpha_p$ the sparse model coefficients, i.e.~$\alpha_p=0$ for most $p$. Following \citet{eckert16}, for $\{\Phi_p\}$ we use a collection of King functions, 

\begin{equation} \Phi_p(R)=\left(1+\frac{R^2}{R_{c,p}^2}\right)^{-3\beta_p}.\end{equation}

King functions have the good property that they are monotonously decreasing and the correspondence between projected (2D) and deprojected (3D) profiles can be written analytically. However, the method proposed here can be readily generalized to any suitable basis. We set up a grid of values of $R_{c,p}$ and $\beta_p$ adaptively to provide a range of shapes that is as general as possible. For a profile with $N$ data points with radii between $R_0$ and $R_{\max}$ we set $N/4$ values of ${R_c}$ logarithmically spaced between $R_0$ and $R_{\max}/2$, and we draw 10 values of $\beta_p$ linearly in the range $0.6-3$. Every combination of $r_c$ and $\beta$ is considered, thus our final dictionary contains $5N/2$ functions. 

In observed (projected) space, the basis of functions can be readily translated into the functions $\phi_p(r)$ as 

\begin{equation}\phi_p(r)=\sum_{p=1}^P \left(1+\frac{r^2}{R_{c,p}^2}\right)^{-3\beta_p+1/2},\end{equation}

with $r$ the projected radius. The coefficients $\gamma_p$ of $\phi_p(r)$ are analytically related to $\alpha_p$ as \citep[see Appendix B of][]{eckert16}

\begin{equation}\gamma_p=\frac{\Gamma(3\beta_p-1/2)\sqrt{\pi}R_{c,p}}{\Gamma(3\beta_p)}\alpha_p\label{eq:deproj}\end{equation}
\noindent with $\Gamma$ the Legendre gamma function. We thus decompose the observed projected profile onto the basis $\{\phi_p\}_{p=1}^P$ and fit for $\{\gamma_p\}_{p=1}^P$. 

\subsection{Likelihood function}

In a photon counting X-ray instrument such as \ero, the observed number of counts $N_c$ in a given annulus is a Poisson realization of the expectation value $\mu$. The expectation value in annulus $i$ can be written as

\begin{equation}\mu_i={\rm PSF}\otimes\left(A_{\rm region}t_{\rm eff}\sum_{p=1}^P\gamma_p\phi_p(r_i)\right)+B_i,\label{eq:mui}\end{equation}

\noindent with $A_{\rm region}, t_{\rm eff}$ the area of the annulus and the effective (vignetting corrected) exposure time in the annulus, and $B_i$ the background expectation. Here $\otimes$ denotes the convolution. The best fitting brightness profile will be described by the set of coefficients $\{\hat\gamma_p\}_{p=1}^P$ that maximizes the likelihood function \citep{cash79}

\begin{equation}-\log\mathcal{L}=\sum_{i=1}^N \mu_i-N_{c,i}\log\mu_i. \end{equation}

To enforce a sparse representation on the basis of functions, the likelihood is modified by the modulus of the sum of the coefficients \citep[the \emph{lasso} method,][]{tibshirani96}, such that the estimator $\{\hat\gamma_p\}_{p=1}^P$ can be written as

\begin{equation}
\mathbf{\hat\gamma}=\arg\min\left\{-\log\mathcal{L}(\gamma_p)+\lambda\sum_{p=1}^P\lvert\gamma_p\rvert\right\}\label{eq:lasso}
\end{equation}

The addition of the lasso penalty term penalizes against complicated linear combinations of basis functions, thus ensuring a smooth and simple description of the brightness profile. The value of the thresholding parameter $\lambda$ is set once and for all using Monte Carlo simulations of a smooth curve and selecting the smallest value of $\lambda$ that reproduces the true curve within a given fractional precision, which we choose to be 1\% \citep[see][for details]{diaz17}. Therefore, our model is tuned to reproduce generic smooth curves with a typical precision of 1\%, which given the properties of the \ero\ all-sky survey should always be smaller than the statistical uncertainties. 

In the 2D method presented in \citet{diaz17} point sources are detected and modeled on-the-fly by adding an additional component to the model in the form of a list of sources following a sparse representation across the image. In the faster 1D method proposed here this approach cannot be implemented directly, thus we apply a standard masking of the point sources detected through an external tool.

\subsection{PSF convolution}
\label{sec:psf}

To convolve the 1D model with the PSF, for a pre-defined grid of radii $r_i$ (usually defined as the emission-weighted mean of each annulus) we construct a PSF convolution matrix following \citet{croston06,eckert16}. Assuming that the surface brightness distribution within each bin is roughly constant, which is a good approximation if the radial binning is fine enough, we define PSF$_{i,j}$ as the fraction of photons originating from bin $i$ and measured in bin $j$. To construct the matrix, we use the model for the \ero\ survey PSF presented in \citet[see their Fig. 6]{clerc18}. The total PSF mixing matrix is defined as the mean of all the pixel integrals, i.e.

\begin{equation}
{\rm PSF}_{i,j}=\frac{1}{N_{pix,i}}\sum_{k\in i} \sum_{l\in j}{\rm PSF}(\lvert k-l\rvert),
\end{equation}
\noindent where PSF$(r)$ denotes the normalized functional form for the PSF, the sums are performed over all the pixels in annuli $i$ and $j$, $\lvert k-l\rvert$ is the distance between pixels $k$ and $l$, and $N_{pix,i}$ is the number of pixels in bin $i$. 

In practice, performing such a double integral can be time consuming when the considered image is large. To speed up the computation, we perform the calculation in Fourier space using fast Fourier transforms (FFT). For each radial bin, we create an image with a flat brightness distribution within the corresponding annulus and filled with zeros elsewhere. We convolve the resulting image with the PSF kernel by multiplying the Fourier transformed images and then transforming back to real space. The corresponding row of the PSF matrix is then filled by counting the (normalized) fraction of the flux in the convolved image that falls within each annulus.

\begin{figure}
    \resizebox{\hsize}{!}{\includegraphics{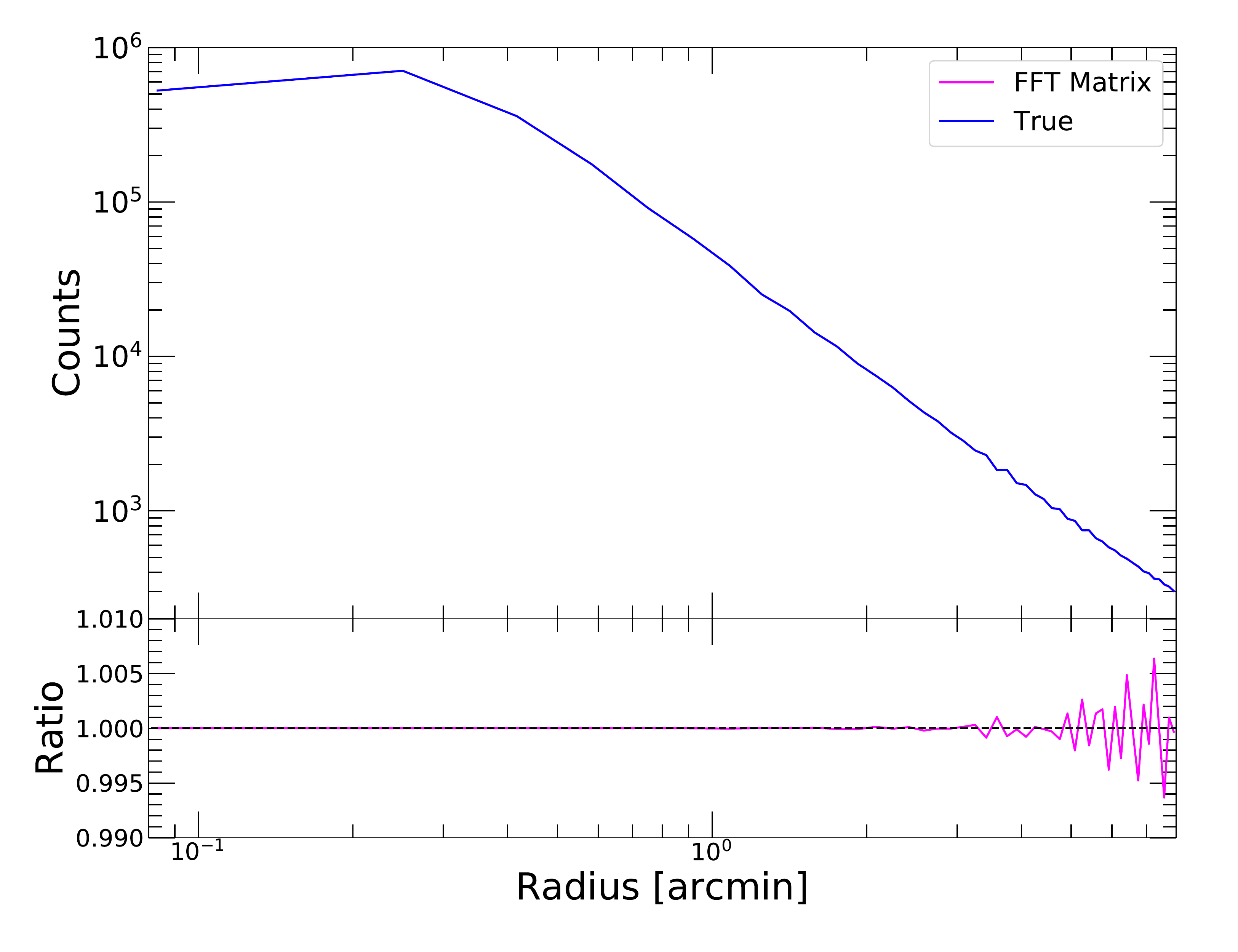}}
    \caption{Accuracy of the PSF convolution matrix calculated with FFT. Comparison between the result of our 1D matrix product computed with FFT (magenta) and direct 2D PSF convolution (blue curve) for a simulated bright point source. The bottom panel shows the relative deviations of the 1D method compared to the computationally expensive 2D convolution.}
    \label{fig:psf_accuracy}
\end{figure}

To verify the accuracy of the computation, we generated an image of a central point source and convolved it with the PSF. We then calculated the radial profile of the convolved point source. On the other hand, we generated the true count profile for the source (corresponding to a delta function) and multiplied it with the PSF matrix in 1D. In Fig. \ref{fig:psf_accuracy} we compare the 1D convolution with the 2D one. The 1D convolution method, albeit much simpler, reproduces the results expected from the 2D convolution within less than 1\% out to 8 arcmin from the core of the PSF, where the flux of the wings is over 3 orders of magnitude smaller than in the core. Thus, we conclude that for our purposes performing the PSF convolution in 1D is largely sufficient.

Another advantage of our approach is that once the PSF matrix is computed, the calculation of the expectation value $\mu_i$ (see Eq.~\ref{eq:mui}) can be written as a linear function of the coefficients $\{\gamma_p\}$. Once the radial binning is set, the basis functions can be evaluated once and for all on the pre-defined grid, and we can construct a $N\times P$ convolution matrix $K$ including all the terms in Eq.~\ref{eq:mui} such that $\mu$ becomes the matrix product of the vector of coefficients $\mathbf{\gamma}$ with the matrix $K$. Thus, the problem reduces to solving a generalized linear model.


\subsection{Optimization}
\label{sec:optim}

To solve the lasso optimization problem (Eq.~\ref{eq:lasso}), we implement the model (Eq.~\ref{eq:mui}) as a generalized linear model within the probabilistic programming framework in the Python package \texttt{PyMC3} \citep{pymc3}. We use the No-U-Turn Sampler \citep[NUTS,][]{nuts} as implemented in \texttt{PyMC3} to sample the posterior distributions of the parameters. NUTS is a Hamiltonian Monte Carlo sampler that makes use of gradient information for fast convergence. This is required to sample efficiently high-dimensional problems ($P>N$) such as the case presented here. In addition, NUTS includes several automatic ways of tuning the Hamiltonian Monte Carlo sampler to improve the sampling efficiency without requiring any user intervention, which makes it suitable for use on large datasets.

We first run a traditional maximum likelihood optimization and use the resulting parameter set as a starting point for NUTS. We set broad log-normal priors on the parameter values around the maximum likelihood results to enforce positivity and set parameter boundaries, whilst still leaving a high level of freedom to the sampler. We then draw 1,000 posterior samples using NUTS. 

In case the expectation value of the background $B_i$ is known, the corresponding parameter in Eq.~\ref{eq:mui} can be fixed. Generally speaking, the background value is only known within some range. The uncertainty in the background can be propagated to the posterior distributions by adding it as one or more additional model parameters, on which priors can be set according to the level of accuracy on the background expectation. For the purpose of this paper, we assume that the background is constant across the field of interest and that its true value is known with an accuracy of 5\%, which is similar to what can be achieved with \emph{XMM-Newton} \citep[e.g.][]{ghirardini17}. Thus, we include one additional model parameter to describe the background, and we set a Gaussian prior on the background parameter with a mean centered on the expected value and a standard deviation of 5\%. The method can be easily generalized in the case of spatially varying background by supplying a background map, whose value will be used as a prior for the expectation values of $B_i$ in Eq.~\ref{eq:mui}.

\begin{figure}
	\resizebox{\hsize}{!}{\includegraphics{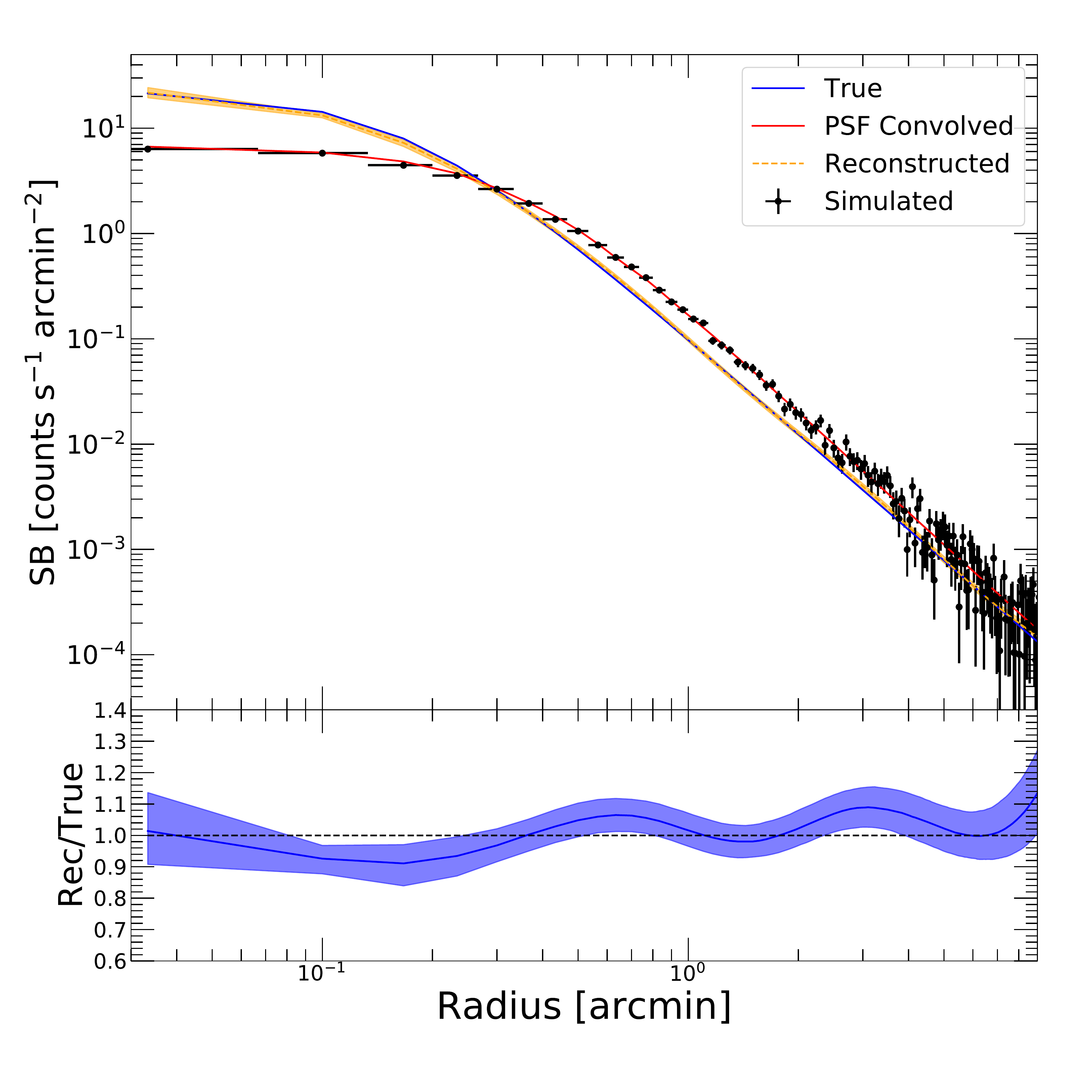}}
	\caption{Example reconstruction for a high-S/N simulated cluster described by a beta-model with $r_c=10^{\prime\prime}$. The true cluster image(blue curve) was convolved with the \ero~PSF (red). The black data points show a Poisson realization of the PSF-convolved image, from which the profile was reconstructed (orange curve). The bottom panel shows the ratio between the reconstructed and the input profile. }
	\label{fig:highnc}
\end{figure}

In Fig. \ref{fig:highnc} we show a test of profile reconstruction on a simulated profile of a compact source with known background and high statistics (10,000 input source counts). From the true profile, we created a spherically symmetric image of the source, which was then convolved in 2D with the \ero~PSF. A Poisson realization of the convolved image was generated, and our reconstruction and optimization technique was applied to the simulated data. The comparison between the reconstructed and true profiles (bottom panel) shows that our technique is able to reproduce the true profile shape all the way to the center with differences of at most 5\% in case the true PSF and background is known.

\subsection{Density profile and mass proxy estimation}

Once the optimization has been performed and the output samples of the parameters $\mathbf{\gamma}$ have been estimated, the projected and 3D profiles can be estimated using Eq.~\ref{eq:3D}, given the analytical conversion Eq.~\ref{eq:deproj}. The conversion between count rate and emissivity can be estimated in a standard way by computing the cooling function $\Lambda(T,Z)$ in the energy band of interest with a plasma emission code such as APEC \citep{apec} or SPEX \citep{spex}. For soft X-ray bands (e.g. $[0.5-2]$ keV), the cooling function is nearly independent of temperature and metallicity as long as the gas temperature exceeds $\sim3$ keV. See for instance \citet{eckert16} for a detailed description of the process. 

The core-excised luminosity in the radial range $[r_1,r_2]$ can then be evaluated by integrating the output profile,

\begin{equation}L_{X,ce}=\int_{r_1}^{r_2}\Lambda(T,Z)n_e(r)n_p(r)\,2\pi r\,dr\end{equation}

\noindent with $n_e(r), n_p(r)$ the number densities of electrons and protons, respectively. The emission measure is proportional to the surface brightness with a proportionality constant $C$ that depends on the telescope's effective area. Given that $EM(r)$ is written as a linear combination of basis functions (Eq.~\ref{eq:3D}), $L_{X,ce}$ can itself be written as a linear combination with coefficients $L_p$ proportional to $\gamma_p$. The integral of the basis functions $\{\phi_p\}$ can be written analytically as

\begin{equation} L_p:=\int_{r_1}^{r_2}\phi_p(r)\, 2\pi r\, dr = \frac{2\pi R_{c,p}^2}{3-6\beta_p}\left[\left(1+ \frac{r^2}{R_{c,p}^2}\right)^{-3\beta_p+3/2}\right]_{r_1}^{r_2}\end{equation}

The posterior distribution of $L_{X,ce}$ can then be drawn from the posterior samples of $\{\gamma_p\}$ as

\begin{equation}
L_{X,ce}=\Lambda(T,Z)C \sum_{i=1}^P\gamma_p L_p.
\end{equation}

The gas density profile can be reconstructed by converting the 2D emissivity profile into 3D (Eq.~\ref{eq:3D}), 

\begin{equation}
	n_e(r)=\left(\Lambda(T,Z)C \mu_{ep}\sum_{i=1}^P \alpha_p\Phi_p(r)\right)^{1/2},
\label{eq:gasdens}
\end{equation}

\noindent with $\alpha_p$, $\Phi_p$ as defined in Sect.~\ref{sec:model} and $\mu_{ep}=n_{e}/n_{p}\approx1.17$ the ratio of the number density of electrons to that of protons in a fully ionized plasma. The posterior distribution of $M_{\rm gas}$ can also be obtained by integrating Eq.~\ref{eq:gasdens} over the volume \citep[for details see][]{eckert19}.

\section{Monte Carlo Simulations}

To validate the procedure, we performed two sets of Monte Carlo simulations and quantified the ability of our algorithm to recover the core-excised luminosity, gas mass and gas density profile of galaxy clusters even in the low-count and resolution-limited regime, as will be the case for most sources detected in the course of the \ero~ all-sky survey. First, we start from the simple case of spherical clusters following a beta-model \citep{cavaliere76} and a uniform background. Second, we simulate real morphologies determined from deep X-ray images of a large set of clusters, and include a realistic AGN distribution. In both cases, we use SIXTE \citep{sixte} to generate realistic \ero~simulations including a wide variety of instrumental effects.

\subsection{Beta-model simulations}

As has been done in numerous previous papers \citep[e.g.][]{pacaud06,clerc12,clerc18,kaefer19b}, we start by running synthetic simulations of individual systems with a gas distribution following a single spherically symmetric beta-model \citep{cavaliere76},

\begin{equation}
S_X(r)=S_0 \left( 1+\left(\frac{r}{r_c}\right)^2\right)^{-3\beta+0.5}.
\end{equation}

We fix $\beta$ to the value of $2/3$ usually measured in local clusters \citep[e.g.][]{mohr99,chen07,kaefer19}. We then vary the core radius $r_c$ in the range 10-60 arcsec and the central normalization $S_0$, assuming $R_{500}=5r_c$. We simulate \ero~observations using SIXTE for a single, uniform exposure time of 3 ks similar to the median exposure time of the final all-sky survey and the field-of-view averaged PSF and a uniform background \citep{clerc18}, assuming a constant gas temperature. From the generated event files we extract photon images in the [0.5-2] keV band with a pixel size of $4\times4$ arcsec and the corresponding exposure maps including correction for vignetting. 

\begin{figure}
	\resizebox{\hsize}{!}{\includegraphics{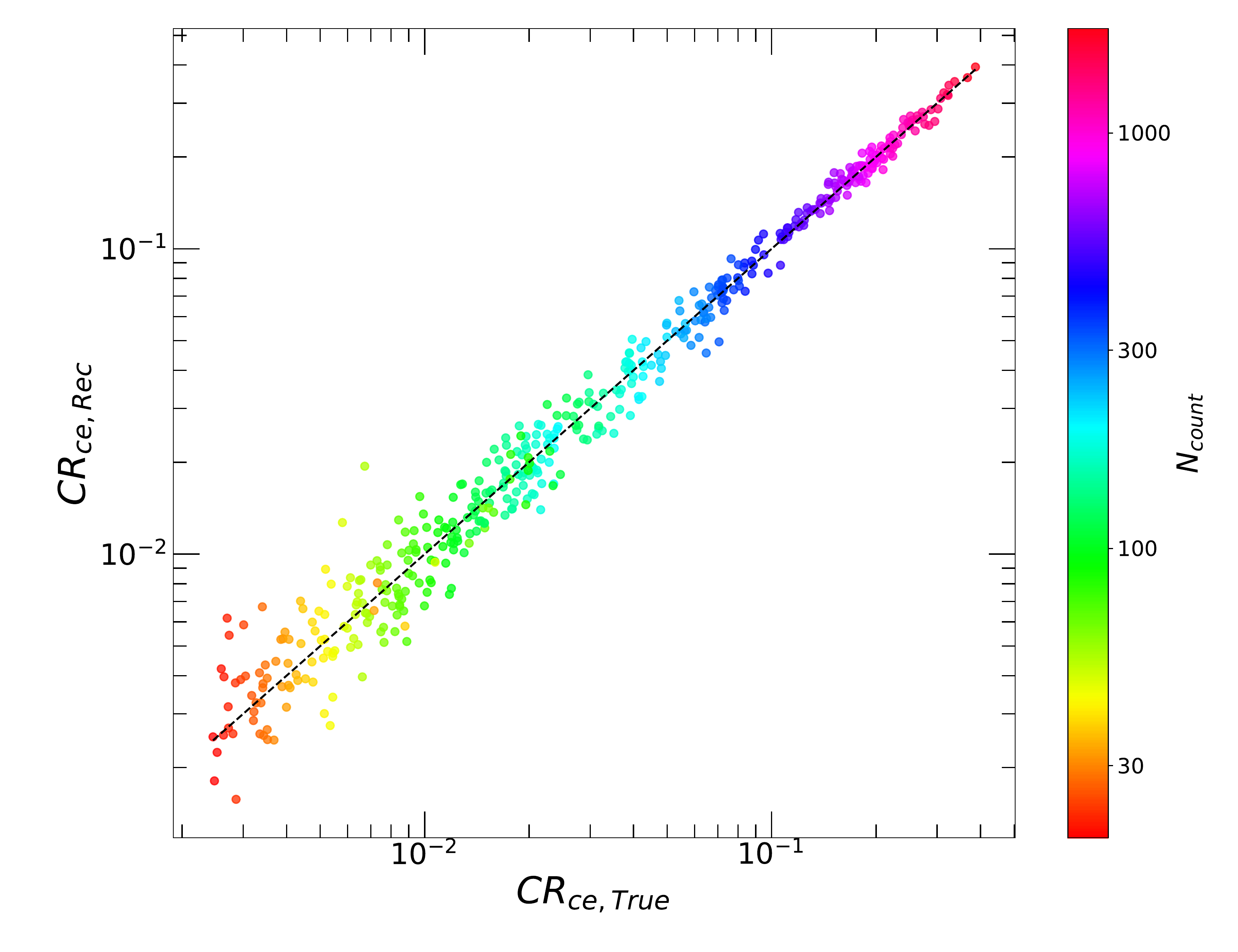}}
	\caption{\label{fig:lxce_rec} Reconstructed core-excised count rate in the $[0.2-0.8] R_{500}$ range plotted against the true value for the beta-model simulation. The data points show the median of the posterior distribution. The data points are color coded by the number of source photons per simulated system (right-hand color bar).}
\end{figure}

\subsubsection{Core-excised luminosity estimation}

We accumulate the photon counts from the simulated maps in circular annuli at maximum resolution (4 arcsec). We reconstruct the profile following the procedure devised in Sect.~\ref{sec:method}. We extract the profiles out to 10 arcmin from the cluster center, i.e.~always well beyond $R_{500}$. This allows a robust determination of the background level outside of the cluster. The parameters describing the source and the background are fitted jointly, such that the output profile is marginalized over the uncertainty in the local background level. We model the background as a constant in Eq.~\ref{eq:mui} and set a Gaussian prior centered on the true background value and a standard deviation $\sigma=0.05$, i.e.~we assume that the local background value can be determined \emph{a priori} with an accuracy of 5\%. We build a PSF mixing matrix using the method described in Sect.~\ref{sec:psf} and then optimize the problem using the NUTS algorithm as described in Sect.~\ref{sec:optim}. We then compute the integrated core-excised count rate for each step in the output chains, and determine the median and 68\% confidence interval from the posterior distribution of values. 

In Fig. \ref{fig:lxce_rec} we show the comparison between the true and the reconstructed core-excised count rates in the $[0.2-0.8]R_{500}$ radial range for 1,000 beta-model simulations. Each point is the median of the posterior distribution for a single realization. For clarity, the results are color coded by the number of total source counts in the simulated profiles. We can see that the reconstructed count rates closely follow the input parameters, even in the low count rate regime. The dispersion of the points around the one-to-one line decreases with increasing number of counts, as expected for purely statistical uncertainties.

\begin{figure*}
	\resizebox{\hsize}{!}{\includegraphics[width=0.5\textwidth]{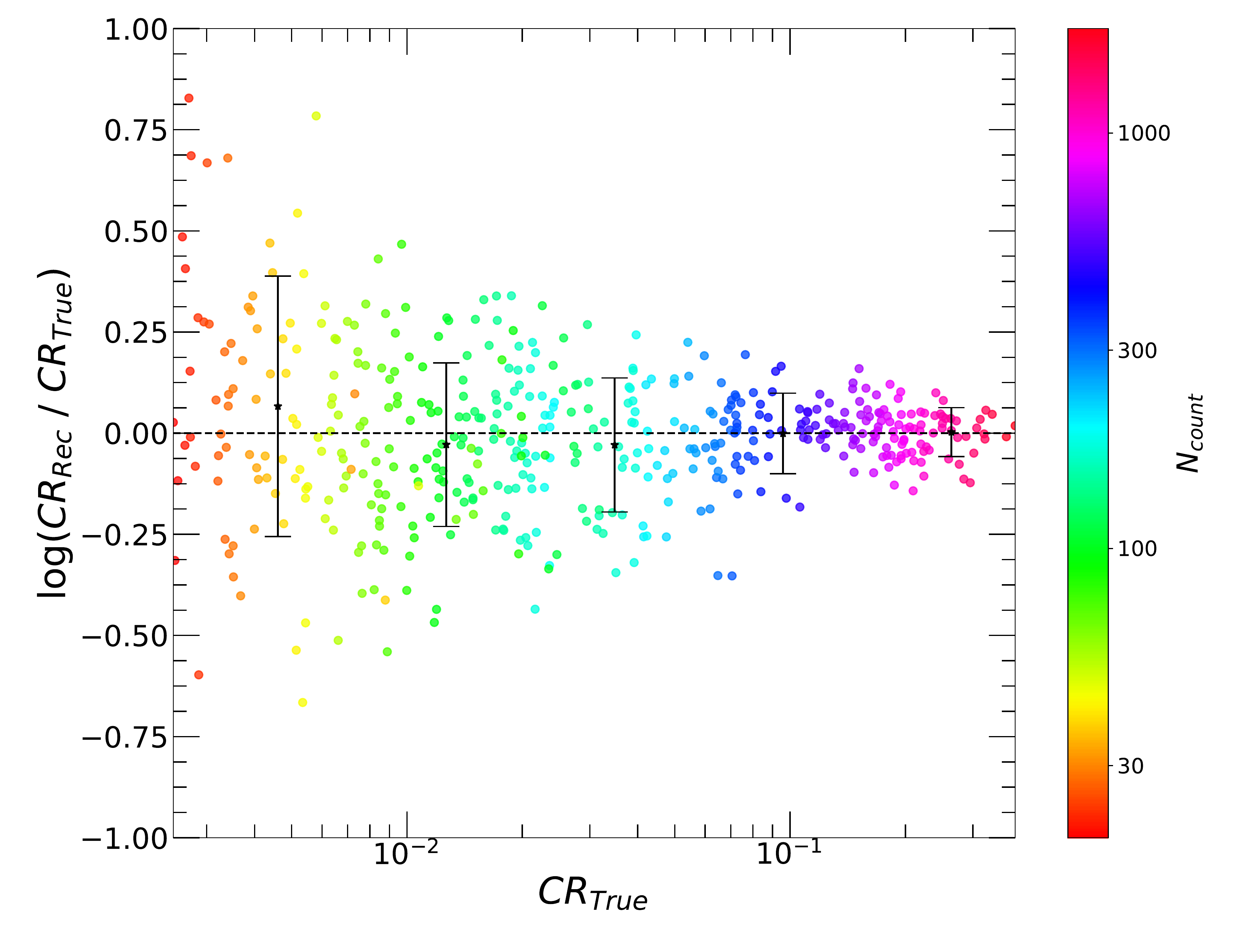}\includegraphics[width=0.5\textwidth]{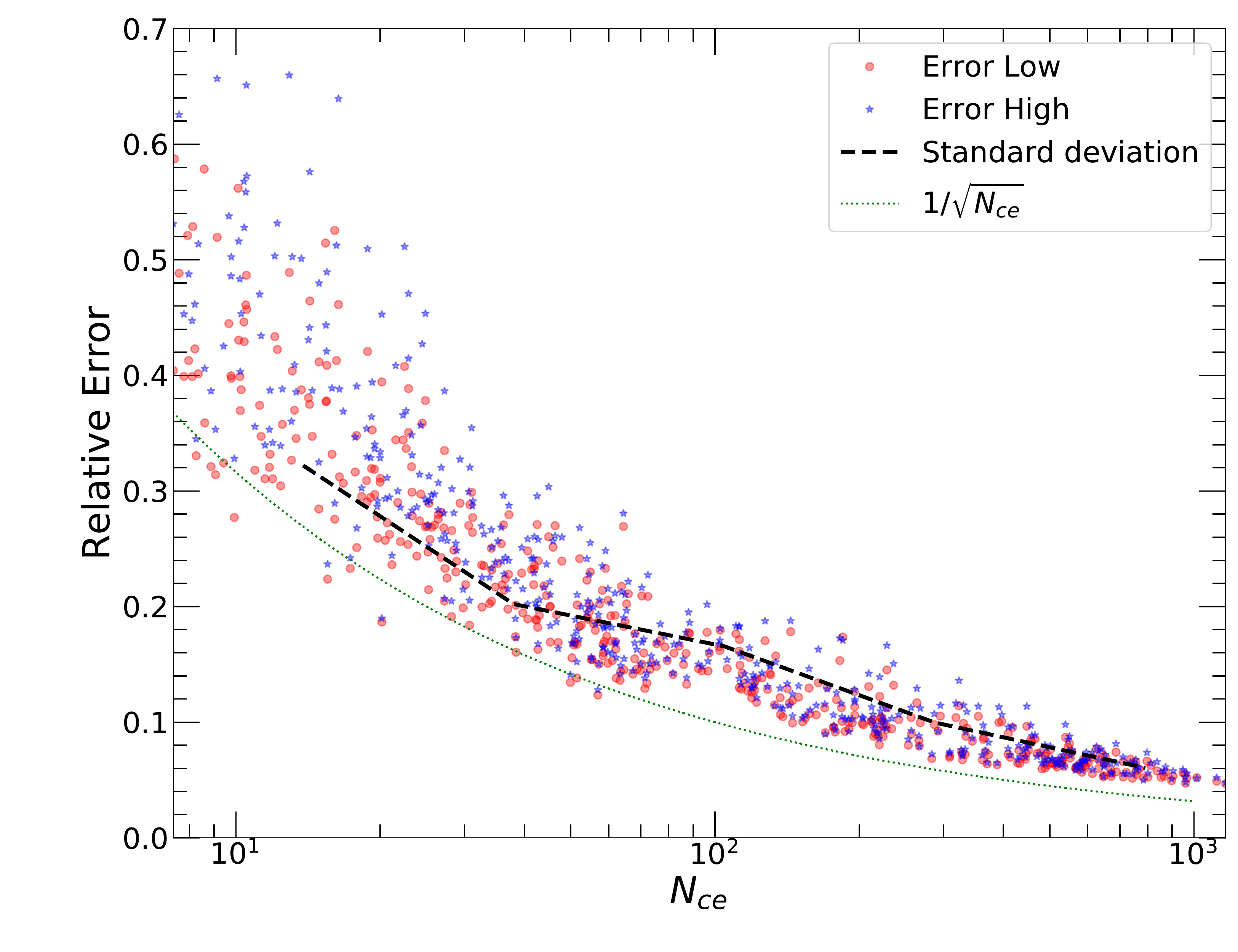}}
	\caption{Bias and scatter of the reconstructed core-excised count rates for the beta-model simulations. The left-hand panel shows the difference in natural logarithm between the true and the reconstructed values as a function of the input source count rate, color-coded by the number of true source counts. The black data points show the median and $1-\sigma$ percentiles of the point estimates in bins of count rate. The right-hand panel shows the relative uncertainty of the reconstructed core-excised count rates as a function of the true number of core-excised source counts. The points show the lower (red) and upper (blue) statistical uncertainties estimated by our algorithm. The dashed black curve shows the standard deviation of the point estimates in bins of count rate (black points in the left-hand panel). For comparison, the green dotted line shows the $1/\sqrt{N_{ce}}$ curve corresponding to the expectation of pure Poisson statistics.}
	\label{fig:errors}
\end{figure*}

\subsubsection{Bias and uncertainties}

The idealized beta-model simulations can also be used to calibrate the level of bias introduced by the reconstruction technique and to test the accuracy of the reconstructed error bars. In the left-hand panel of Fig. \ref{fig:errors} we show the deviations from the one-to-one relation, defined as the log of the ratio of true to reconstructed count rates. We group the data in bins of true count rate, and measure the median and dispersion of the points in each subset. Our procedure is able to measure the core-excised count rate in the idealized simulations with no measurable bias (always better than 5\%) and a dispersion that closely follows the input count rate.

In the right-hand panel we compare the error bars determined by our algorithm (split into upper and lower 1-$\sigma$ uncertainties) with the dispersion of the points around the median, as determined in the left-hand panel of the figure. The data points are plotted against the number of true core-excised counts $N_{ce}$ ($[0.2-0.8]R_{500}$) of each realization. We find that the error bars calculated from the posterior distributions closely match the dispersion of the data points, indicating that our error bars are accurate. For comparison, we also show the expectation from pure Poisson statistics, in which case the variance should scale as $N_{ce}$. We can see that the error bars in the reconstructed count rate always exceed the Poisson expectation, which is unsurprising given that deconvolution from the PSF introduces an additional level of complexity. The error bars obtained with our method are increased by $\sim30\%$ with respect to the case of an ideal PSF. Therefore, we conclude that our deprojection and PSF deconvolution method is able to measure accurately the core-excised flux of \ero~ clusters even in the low-count regime, albeit with a moderate increase in the statistical errors.

In practice, \ero~ is not expected to be able to distinguish extended sources from point-like sources for less than $\sim40$ source counts \citep{pillepich18,grandis18,kaefer19b}. The threshold of 40 counts corresponds to a relative uncertainty of $\sim30\%$ on the core-excised count rate. This number should be put in perspective of the expected scatter in the $L_X-M$ relation. For instance, \citet{maughan12} measure a scatter 67\% for the $L_X-T$ relation, which decreases to 29\% when excising the central regions ($R<0.15R_{500}$). The scatter may be decreased even further when excising the core out to $0.2R_{500}$, given that the median self-similar scaled profiles of the CC and NCC populations are indistinguishable beyond that radius \citep{kaefer19,ghirardini19}. \citet{mantz18} suggest that the scatter of the relation between core-excised $L_X$ and $M_{500}$ could be as low as 10\% for massive clusters. Therefore, the lower scatter of the $L_{X,ce}-M$ relation more than makes up for the modest increase in statistical uncertainties compared to the total integrated $L_X$.

\subsection{Mock \emph{eROSITA} field}

To extend our tests beyond the idealized case of a single beta-model, we apply our technique to a simulated field tuned to reproduce the expected properties of the \emph{eROSITA Final Equatorial Depth} (eFEDS) survey, which was observed by \ero~during the performance verification (PV) phase. eFEDS targets a 120 deg$^2$ field at the final depth of the \ero~ survey and serves as a test bed for the expected performance of the instrument. Most of the survey area lies within the footprint of the Subaru HSC-SSP survey \citep{miyazaki18}, which will readily provide an optical identification of the detected clusters, precise photometric redshift estimates thanks to the multi-color nature of HSC-SSP, and weak lensing mass calibration. The clusters in the simulated eFEDS field are drawn from real X-ray images of known clusters, thus the simulation encompasses a wide variety of radial shapes and morphologies. Here we assess the performance of our algorithm on the simulated eFEDS field.

\subsubsection{The eFEDS mock}

\label{sec:rec}
\begin{figure*}
	\resizebox{\hsize}{!}{\includegraphics[width=0.5\textwidth]{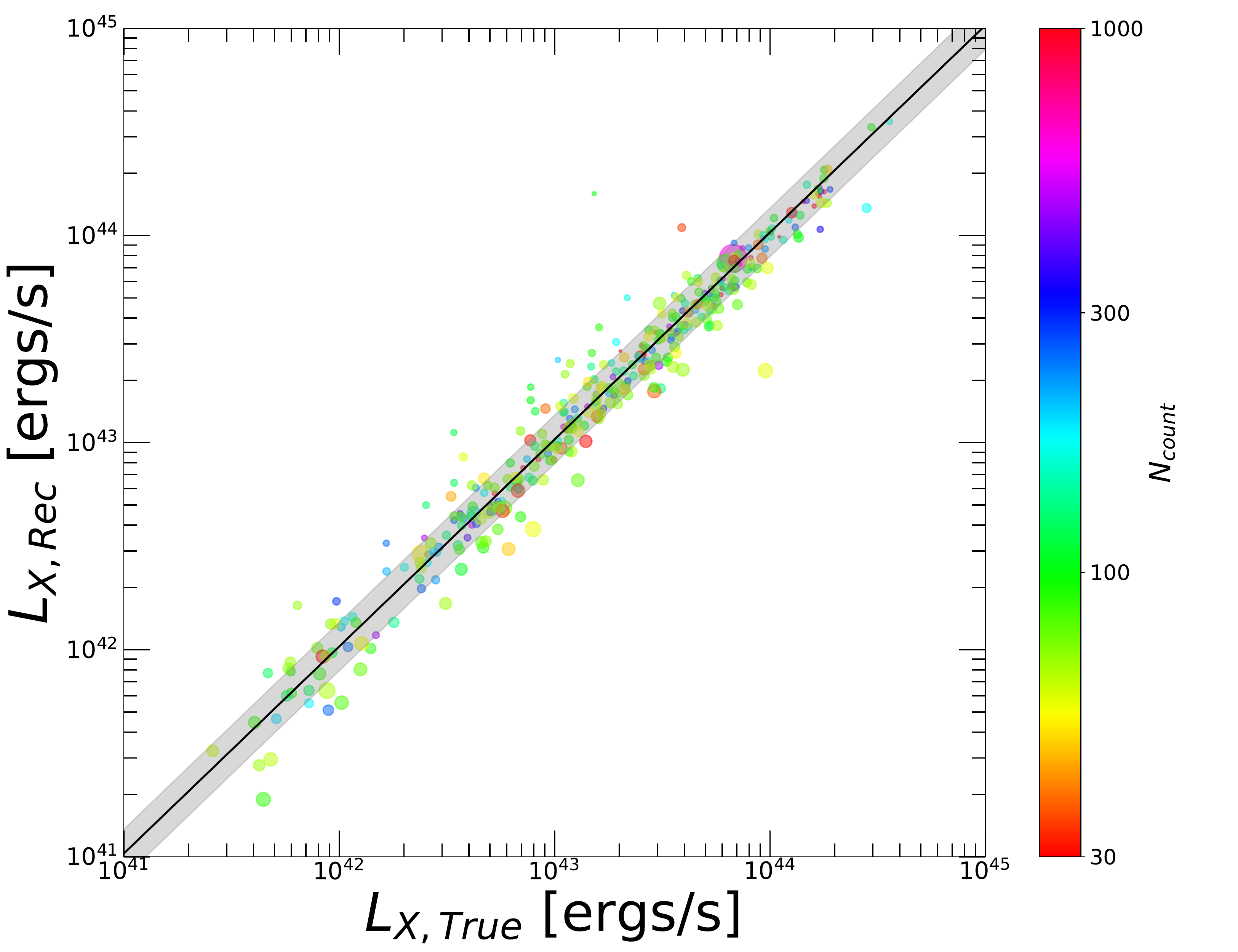}\includegraphics[width=0.5\textwidth]{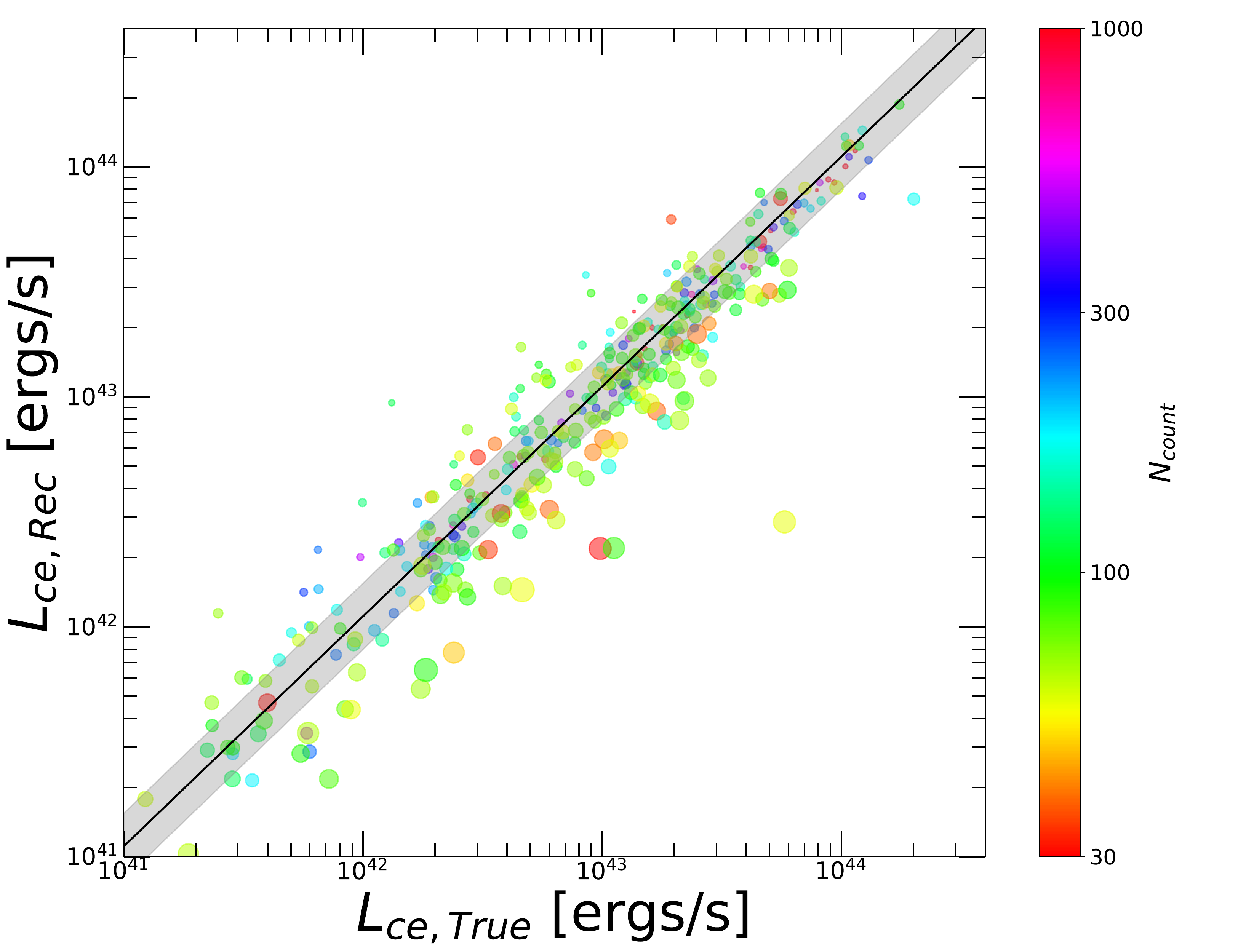}}
	\caption{Reconstructed luminosity against true luminosity for the case of the full luminosity (left) and the core-excised luminosity ($[0.2-1]R_{500}$, right). The points are color-coded by the number of reconstructed source counts (color bars) and the symbol size is proportional to the uncertainty. The black lines and shaded areas show the best-fit power law relations and the fitted intrinsic scatter.}
	\label{fig:efeds_lx}
\end{figure*}

The creation of the mock cluster sample follows the procedure outlined in \citet{grandis18}. We draw a cluster catalogue with halo masses $M_{500c}$ and redshifts from the \citet{tinker08} halo mass function and distribute the halos uniformly over the eFEDS footprint. For each cluster, we then assign luminosities in the rest frame $[0.5-2]$ keV band and temperatures using the scaling relation and the scatter around that relation reported by \citet{bulbul19} for 59 \emph{XMM-Newton} follow up observations of SPT-SZ selected clusters. We assume that the scatter in luminosity and the scatter in temperature are uncorrelated. The metallicities of the clusters are set to $Z=0.3\,Z_\odot$ \citep{mcdonald16}. We refer to \citet{grandis18} for more details on the mock catalogue creation.

The cluster spectrum is predicted with an APEC model, using the aforementioned temperature and metallicity, as well as the neutral hydrogen column density at the cluster position. The cluster 2D surface brightness is simulated by randomly selecting one of 83 \emph{Chandra} images of SPT-SZ selected clusters \citep{sanders18} or 37 \emph{XMM-Newton} images of nearby lower mass ($0.7$ to $7\times 10^{14}\,\mathrm{M}_\odot$) objects drawn from the eeHIFLUGCS sample \citep{ramos19} which fit within the \emph{XMM-Newton} field of view. Masses for these input systems were assumed to be the value from the SPT catalog or using \emph{XMM-Newton} \citep{piffaretti11}, as appropriate. The images of these clusters were adaptively smoothed to have a signal to noise ratio of 8--10 within the smoothing kernel, also subtracting background. As the input images for the simulation had to be zero or positive, negative values in the images were replaced by zeros and a tapering was applied to the cluster image at larger radii to help remove sharp edges and counteract the additional flux redistribution towards the outskirts that the zeroing would create.

For a mock cluster with a mass of greater than $10^{14.55}\,\mathrm{M}_\odot$, a random input image was taken from those real clusters with masses greater than that threshold in an appropriate redshift bin ($<0.4$, $0.4$--$0.6$, $0.6$--$0.8$, $0.8$--$1.0$ and $>1.0$). For mock cluster masses lower than this, a random image of a cluster in mass bins of $10^{14.40}$--$10^{14.55}$, $10^{14.22}$--$10^{14.40}$ and $<10^{14.22}\,\mathrm{M}_\odot$ was chosen. The image was scaled in size by the ratio of $R_{500c}$ of the catalog cluster in arcmin compared to the real cluster and randomly rotated. The flux in the simulation was set to be that in the mock catalog. Correlations between the morphological properties of the clusters and the scatter in luminosity and temperature are neglected. 

Active galactic nuclei (AGN) are added to the mock catalog following the approach presented in \citet{comparat19}. The position and brightness of simulated AGN follows the expected luminosity function and halo occupation distribution of the AGN population, such that the simulated point-like sources follow realistic brightness and clustering properties.

\subsubsection{Reconstruction}

The full eFEDS field was simulated with SIXTE using the same attitude file as for the actual survey. The simulated data were processed using the standard \ero~ software package (eSASS). From the simulated event files, we extracted mock \ero~ images in the [0.5-2] keV band combining all 7 telescope modules. A merged exposure map was computed using the eSASS task \texttt{expmap}. We used the standard eSASS detection algorithm \citep{clerc18} to detect point-like sources in the field and prepared a point-source mask to excise circular regions around each point source, with an exclusion radius that is proportional to the logarithm of the detected count rate and a minimum radius of 15 arcsec.

From the original eFEDS mock catalogue, we performed a simple cut on the true, total flux and selected all the halos with $f_X>3\times10^{-14}$ erg/s/cm$^2$, which yields a total of 445 simulated sources. In each case, we masked the surrounding point sources and extracted the count profile in circular annuli, as was done for the idealized beta-model simulations. We fixed the redshift of each cluster to the catalog value and ran the reconstruction of the surface brightness profiles, X-ray luminosity (total and core excised), gas density profile, and integrated gas mass. 

In Fig. \ref{fig:efeds_lx} we show the result of the reconstruction of the X-ray luminosity within $R_{500}$. As a first step, we assume that the value of $R_{500}$ is known and fixed to the catalog value. We show the results both for full, integrated luminosity (left-hand panel) and for the core-excised luminosity in the $[0.2-1]R_{500}$ range (right-hand panel). The data points are color coded by the number of reconstructed net source photons. We can see that the both the total and the core-excised luminosities are recovered very accurately by our method, down to the detection threshold of $\sim50$ counts. Given that the sample selection was performed on the true simulated flux, a number of systems actually exhibit a very low number of counts ($<50$) because their exposure time is low. These systems would not be detected in the real survey. While their recovered fluxes are not necessarily biased (see Fig. \ref{fig:errors}), they exhibit a large statistical scatter. Thus, in Fig. \ref{fig:efeds_lx} we display only the clusters for which the true number of source counts is at least 30, which still encompasses all the detected systems. Fitting the relations with a power law with free intrinsic scatter, we find that the luminosities are recovered with no measurable bias and an intrinsic scatter $\sigma_{\ln L_{X,tot}}=0.12\pm0.02$ and $\sigma_{\ln L_{X,ce}}=0.21\pm0.04$. Therefore, excising the central regions moderately increases the scatter in the reconstruction of the true cluster $L_X$. 
 
Around the detection limit the relative uncertainty in $L_{X,ce}$ is about 30\% and it decreases for brighter sources. For the entire population of 445 sources, the relative uncertainty can be approximately described by a power law with an index of $-0.5$ as expected in the case of Poisson statistics,

\begin{equation}\label{eq:obs_err_lceiter}
    \frac{\Delta L_{X,ce}}{L_{X,ce}} \approx 0.33 \left(\frac{N_{c}}{50}\right)^{-0.48}.
\end{equation}

From the reconstructed profiles we can readily reconstruct the gas density profiles (Eq.~\ref{eq:gasdens}) and the gas mass by integrating the gas density over the volume. Since the gas density is a slow function of the X-ray surface brightness, the posterior uncertainties in $M_{\rm gas}$ are smaller than on $L_{X,ce}$. Again approximating the error budget for $M_{\rm gas}$ as a power law of the number of source counts, we obtain the relation

\begin{equation}\label{eq:obs_err_mgas}
    \frac{\Delta M_{\rm gas}}{M_{\rm gas}} \approx 0.21 \left(\frac{N_{c}}{50}\right)^{-0.40}
\end{equation}

i.e.~around the detection limit $M_{\rm gas}$ can be reconstructed with an uncertainty of about 20\%. The relative uncertainty in the gas mass is lower than in the core-excised luminosity since the gas density is proportional to the square root of the emissivity. 

\subsubsection{Mass estimation}

Assuming an externally calibrated scaling relation between core-excised $L_X$ and total mass, we can use the core-excised luminosity as a low-scatter proxy to reconstruct the mass of the detected galaxy clusters. To do so, we developed an iterative procedure based on the reconstructed brightness profile deconvolved from PSF effects. Our input catalogue for the eFEDS mock encodes an $L_{X,ce}-M_{500}$ relation which can be described as

\begin{equation}
    L_{X,ce} = 3.6\times10^{42}\left(\frac{M_{500}}{10^{14}M_\odot}\right)^2 E(z)^{7/3} \mbox{ ergs/s}
    \label{eq:lxm}
\end{equation}

with $E(z)=\sqrt{\Omega_m(1+z)^3+\Omega_\Lambda}$ used to correct for the self-similar evolution of the $L_X-M$ relation. The input log-normal intrinsic scatter of the relation is $\sigma_{\ln M}=0.18$ at fixed $L_{X,ce}$. Given the lower statistical uncertainty on $M_{\rm gas}$ compared to $L_{X,ce}$, a similar procedure can be put together with the gas mass as a mass proxy, potentially leading to more accurate results. However, for low cluster temperatures ($T\lesssim 3$ keV) the cooling function $\Lambda(T,Z)$ is a strong function of metal abundance, which is impossible to measure from survey data in most cases. This introduces an additional source of systematic uncertainty in the reconstruction of $M_{\rm gas}$, whereas $L_{X,ce}$ is much less affected by this potential bias. For high-temperature clusters ($T>3$ keV) $M_{\rm gas}$ can potentially be used as a highly effective proxy for the cluster mass.

Assuming that the scaling relation and the cluster redshift are known, we set up an iterative procedure to determine the aperture within which $L_{X,ce}$ should be integrated (i.e.~$R_{500}$ in angular coordinates) and eventually the cluster mass. We start by making a guess for the value of $R_{500}$, which we take to be 700 kpc. We integrate the count rate and then the luminosity within the corresponding aperture, and go through the scaling relation (Eq.~\ref{eq:lxm}) to get a new estimate of $M_{500}$ and hence of $R_{500}$. We repeat the procedure until it converges, i.e.~until the difference of $R_{500}$ values between two iterations is less than 1 kpc. Convergence usually occurs after $\sim5$ iterations.

To validate this procedure, we compared the values of $L_{X,ce}$ computed through this technique with the values recovered inside the true $R_{500}$ aperture as in Sect.~\ref{sec:rec}. We found that our iterative procedure recovers the correct aperture very accurately, with typical differences in $L_{X,ce}$ of less than 5\% and a median ratio $L_{\rm iter}/L_{500}=1.01$. Thus, the uncertainty induced by the unknown aperture of the detected systems is much smaller than the statistical uncertainty and the $L_X-M$ scatter. 

\begin{figure}
\resizebox{\hsize}{!}{\includegraphics[width=0.5\textwidth]{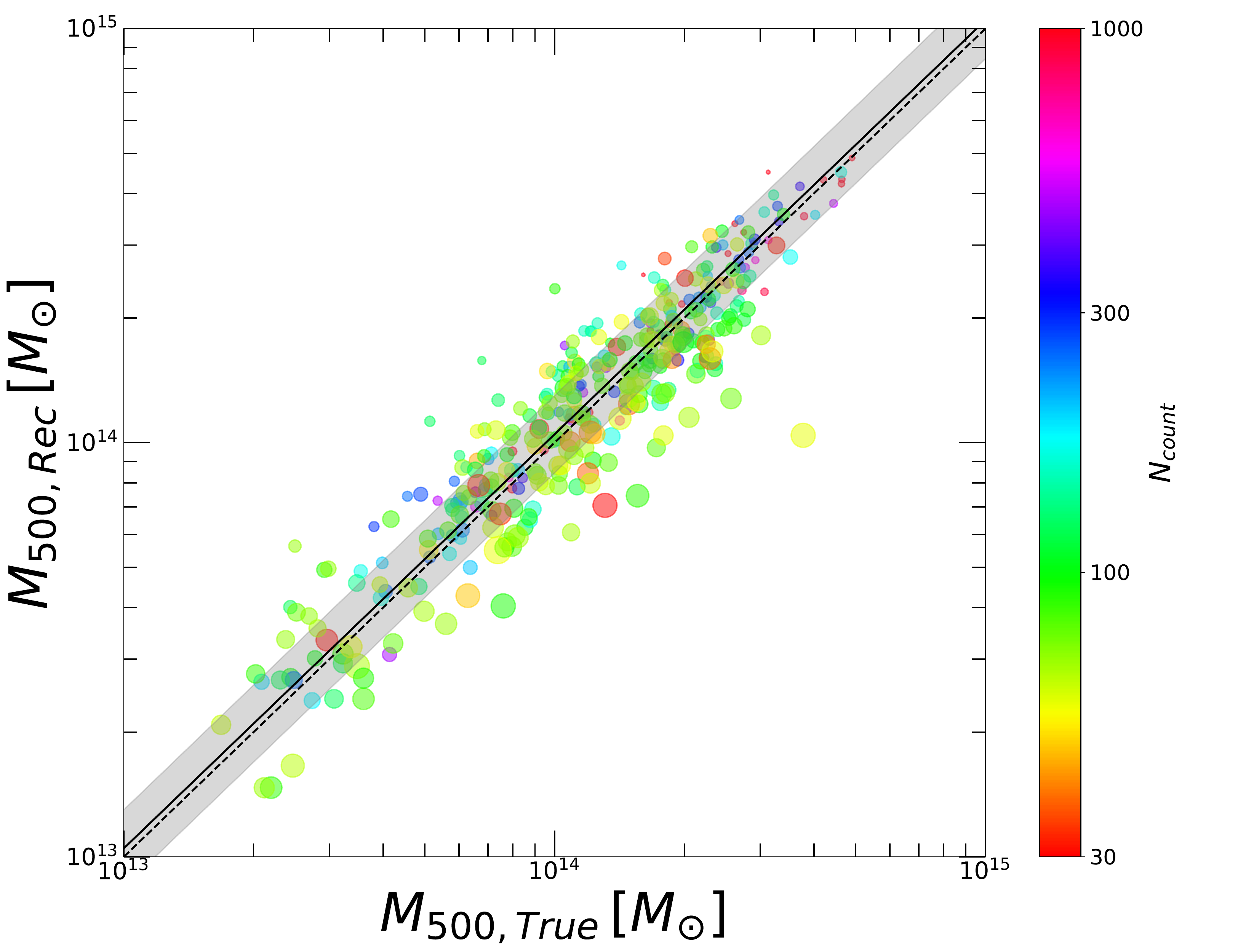}}
\caption{\label{fig:m500} $M_{500}$ values reconstructed through our iterative procedure plotted against the true cluster mass for the eFEDS simulation. The color code shows the number of reconstructed source counts. The black line and shaded area show the best-fit linear relation and its intrinsic scatter, whereas the dashed line gives the one-to-one relation.}
\end{figure}

Finally, we compared the values of $M_{500}$ determined through our iterative procedure with the mass given in the input catalogue. The results are presented in Fig. \ref{fig:m500}, where we show the reconstructed mass as a function of the true mass color coded by the number of source counts for the 374 clusters of the eFEDS mock with $N_{c}>30$. We omit the systems with $N_c<30$ as they have very large uncertainty and are substantially below the detection limit of the survey. We can immediately see that the reconstructed mass traces well the true mass, with a slope and a normalization close to unity. Fitting the relation with a linear function (i.e.~fixing the slope to unity), we measure $M_{500,rec}/M_{500,True}=1.04\pm0.03$ and an intrinsic scatter $\sigma_{\ln M_{500,rec}}=0.21\pm0.02$. This value is only mildly larger than the input intrinsic scatter of the simulation ($\sigma_{\ln M}=0.18$). Therefore, we can conclude that even in the case of a realistic population of simulated clusters spanning a wide variety of shapes and redshifts, the scatter of the output relation at fixed mass is dominated by the intrinsic scatter of the $L_{X,ce}-M_{500}$ relation. 

We note that slope of the $L_{X,ce}-M$ relation is relatively steep. Therefore, deviations from the true value of $L_{X,ce}$ imply smaller deviations from the true $M_{500}$, and the scatter introduced by the reconstruction of $L_{X,ce}$ is subdominant with respect to the intrinsic scatter of the $L_{X,ce}-M_{500}$ relation.

\subsection{Implications on mass estimation}

Our results on the observational scatter in the reconstruction of the cluster's gas mass (Eq.~\ref{eq:obs_err_lceiter} and core-excised luminosities (Eq.~\ref{eq:obs_err_mgas}) indicate that for all object in an \ero~selected cluster sample, low scatter mass proxies can be derived. More specifically, assuming the characteristic limiting photon count of $N_c>50$ for selection by \ero~\citep{pillepich12, pillepich18, clerc18, grandis18}, the observational scatter on the gas mass will be bound to be $<20\%$ for all selected clusters. Such a precision on the derived mass is comparable to what is achieved for the SZ effect. For instance, in the particular case of the SPT survey, the default mass proxy is the detection significance $\zeta$ \citep[for instance][]{vanderlinde10, bleem15, bocquet19}, which is related to, but not strictly equivalent with, the integrated Compton parameter $Y$. Its observational uncertainty is given by a Gaussian with variance 1, which for the typical selection criterion $>5$ yields a fractional observational uncertainty $<20\%$. The intrinsic scatter in mass from SZ derived masses is typically $\sim 0.11$ \citep[see, e.g.][]{bocquet19}, while the typical mass scatter from gas masses is $\sim 0.08$ \citep[see, e.g.][]{bulbul19}. The method presented in this paper thus provides a mass proxy for all \ero~selected clusters with a total scatter (observational + intrinsic) at fixed mass which is comparable to the total mass scatter for the mass proxies of SZ selected cluster samples. X-ray derived masses for \ero~selected clusters will thus be at least as precise as SZ derived masses for SPT selected clusters.

\section{Conclusion}

In this paper, we presented a novel method to reconstruct the morphological and photometric properties of galaxy clusters from X-ray survey data in the low count rate regime. Our method is based on the linear decomposition of the observed profile onto a basis of functions and makes use of the linear nature of the projection kernel to recover the 3D profiles. We correct for PSF effects by convolving the 2D model with a mixing matrix that can be computed numerically for any PSF shape. Our method is readily applicable to the upcoming \ero~survey, which is now ongoing.

We validated our method using large standardized sets of simulations, both in the idealized case (beta-model) and in a more realistic case based on real observed clusters. In both cases, we find that our method can recover the input properties with no evidence for biases down to the survey detection threshold of $\sim50$ counts. This allows us to derive in a robust manner core-excised X-ray luminosity and the gas mass, which are known to be low-scatter proxies of the total mass. Unlike the full integrated X-ray luminosity, they are relatively unaffected by the properties of cluster cores, which are strongly affected by baryonic physics. 

Assuming that an accurate external scaling relation between cluster mass and mass proxies is available, we set up an iterative procedure to determine the total mass from our low-scatter mass proxies. In the specific case of the core-excised luminosity as a mass proxy, our method can recover the true cluster mass with a very modest increase in intrinsic scatter (from 0.18 to 0.21 in our simulation). This is to be contrasted with the scatter of the full integrated $L_X$ at fix mass, which is known to be $\sim0.6$. Therefore, excising cluster cores clearly improves the accuracy of the mass estimates compared to the total integrated flux and renders X-ray mass proxies comparable in precision to SZ mass proxies.

For easy replication of our results, we distribute our code in the form of the public Python package \texttt{pyproffit}\footnote{\href{https://pyproffit.readthedocs.io}{https://pyproffit.readthedocs.io}}, which is available on Github and PyPI.

\begin{acknowledgements}
DE thanks Sylvain Sardy, Jairo Diaz-Rodriguez and Philippe Ganz for helpful discussions on the method. We thank the anonymous referee for their useful comments on the manuscript.
\end{acknowledgements}

\bibliographystyle{aa}
\bibliography{lxce}

\end{document}